\documentclass[aps,prd,letterpaper,showpacs,showkeys,twocolumn,floatfix,nofootinbib,superscriptaddress]{revtex4-1} 

\usepackage{amsmath,amssymb,amsfonts}
\usepackage{bm}
\usepackage{graphicx}
\usepackage{subfigure}

\usepackage[svgnames]{xcolor}

\begin{document}

\preprint{ADP-12-04/T771}

\title{Calculating dihadron fragmentation functions in the Nambu--Jona-Lasinio-jet model}

\author{Andrew~Casey}
\affiliation{CSSM and ARC Centre of Excellence for Particle Physics at the Terascale,\\ 
School of Chemistry and Physics, \\
University of Adelaide, Adelaide SA 5005, Australia
\\ http://www.physics.adelaide.edu.au/cssm
}

\author{Hrayr~H.~Matevosyan}
\affiliation{CSSM and ARC Centre of Excellence for Particle Physics at the Terascale,\\ 
School of Chemistry and Physics, \\
University of Adelaide, Adelaide SA 5005, Australia
\\ http://www.physics.adelaide.edu.au/cssm
}

\author{Anthony~W.~Thomas}
\affiliation{CSSM and ARC Centre of Excellence for Particle Physics at the Terascale,\\ 
School of Chemistry and Physics, \\
University of Adelaide, Adelaide SA 5005, Australia
\\ http://www.physics.adelaide.edu.au/cssm
}

\begin{abstract}
The Nambu--Jona-Lasinio-jet model provides a framework for calculating fragmentation functions without the introduction of \emph{ad hoc} parameters. We develop the Nambu--Jona-Lasinio-jet model to investigate dihadron fragmentation functions (DFFs) of the form $D^{h_1,h_2}_q(z_1,z_2)$. Here we studied DFFs for $q\to \{\pi^+\pi^-\}$, $\{\pi^+K^-\}$ and $\{K^+K^-\}$ with $q=u, d, s$. The driving terms, which represent the probability of one of the hadrons being emitted in the first emission step of the quark-jet hadronization picture, dominate the solutions of the DFFs where either $z_1$ or $z_2$ is large, and $z_1$~($z_2$) is the light-cone momentum fraction of the emitted hadron, $h_1$~($h_2$).  The higher order terms, which represent the probability of neither of the hadrons being emitted in the first emission step of the quark-jet, become more significant as $z_1$~($z_2$) is lowered. Finally, we present a sample result for QCD evolution of DFFs, that significantly modify the model solutions when evolved to typical experimental scale of $4~\mathrm{GeV}^2$. 
\end{abstract}

\maketitle

\section{Introduction}

Deep inelastic scattering (DIS) has proven to be an invaluable source of information about the structure of the nucleon~\cite{Thomas:2001kw}. Initially it provided critical information on the relative distribution of momentum between valence and sea quarks and the gluons. As the experimental capabilities have grown so have our ambitions and over the past decade semi-inclusive deep-inelastic scattering (SIDIS) has helped (along with Drell-Yan) to expand our knowledge of quark flavor structure~\cite{Melnitchouk:1999ft,Beckmann:2002rr,Airapetian:2003ct,Barone:2003jp,Airapetian:2004zf,Airapetian:2007mh,Cloet:2009qs,Alekseev:2010ub,Gross:2012sj}. With several new experimental facilities with 100\% duty factor under construction, SIDIS will become even more important. For example, we may finally be able to pin down the elusive $s-\bar{s}$ asymmetry~\cite{Signal:1987gz,Barone:1999yv,Davidson:2001ji,Cloet:2009qs}. Another area of great current excitement concerns the distribution of the spin of the proton~\cite{Ji:1996ek,Ji:1996nm,Bratt:2010jn,QCDSF:2011aa,Hagler:2011zzb,Thomas:2010zzc,Thomas:2008ga,Myhrer:2007cf,Bass:2009ed,Bass:2011zn,Wakamatsu:2009gx,Wakamatsu:2010qj,Wakamatsu:2010cb,:2008px,Bacchetta:2008xw,Barone:2001sp,Leader:2010gz,Leader:2010rb,Leader:2011za,Leader:2011zz,Vogelsang:2009zz}. There polarized SIDIS is potentially extremely valuable through the study of TMDs~\cite{Ceccopieri:2006wy,Bacchetta:2006tn,Gamberg:2010uw,Gamberg:2010xi,Wollny:2010zp,Anselmino:2011ay,Anselmino:2011ch,Aybat:2011ta,Gamberg:2011my,Bacchetta:2011gx}, which will complement work on GPDs~\cite{Goeke:2001tz,Burkardt:2002hr,Diehl:2003ny,Ji:2004gf,Belitsky:2005qn,Boffi:2007yc,Hagler:2011zzb}.

For these studies to achieve their full potential it is vital that we develop the deepest understanding of the fragmentation functions~\cite{Radici:2011ix}, especially their flavor dependence, and ultimately their dependence on spin and transverse momentum. Fragmentation functions are an important theoretical tool in the investigation of scattering reactions, for example in the separation of the flavor dependence of parton distribution functions~(PDFs). Experimental extractions of fragmentation functions from deep-inelastic scattering data~\cite{Hirai:2007cx,deFlorian:2007aj} have increased theoretical activity in this area. Yet the phenomenological extraction of even favored fragmentation functions suffers from significant uncertainty while the situation for the unfavored is worse. This in turn effects the systematic errors associated with the extraction of the flavor dependence of parton distribution functions through SIDIS. These considerations have led us to develop and study the Nambu--Jona-Lasinio-jet model~\cite{Ito:2009zc,Matevosyan:2010hh,Matevosyan:2011ey,Matevosyan:2011vj}. This model builds on the Field-Feynman quark-jet model (FFQJM)~\cite{Field:1977fa}, by using an effective chiral quark model to provide a framework in which calculations of both quark distribution and fragmentation functions can be performed without introducing \emph{ad hoc} parameters. Pion fragmentation functions in the NJL-jet model were calculated in Ref.~\cite{Ito:2009zc}. The NJL-jet model was then extended to include strange quark contributions and kaon fragmentation functions were obtained~\cite{Matevosyan:2010hh}. Further extensions of the model are the inclusion of vector meson, nucleon and antinucleon fragmentation channels~\cite{Matevosyan:2011ey}, and the inclusion of transverse momentum dependence~\cite{Matevosyan:2011vj}.

Dihadron fragmentation functions (DFFs) represent the probability of producing two hadrons in the decay chain of a fragmenting quark. Some recent work in the area of DFFs include Refs.~\cite{Bacchetta:2006un,Zhou:2011ba}. In Ref.~\cite{Bacchetta:2006un}, parameters for a spectator model are fitted to output from the PYTHIA event generator~\cite{Sjostrand:2000wi} tuned for HERMES~\cite{Liebing:2004us} for dihadron fragmentation functions with a dependence on the sum of the light-cone momentum fractions of the two produced hadrons and their invariant mass squared. DFFs for large invariant mass are studied in Ref.~\cite{Zhou:2011ba}. The dihadron fragmentation functions evolution equations are derived in Ref.~\cite{deFlorian:2003cg} from factorization of the cross-section for the production of two hadrons in $e^+e^-$ annihilation in the $\overline{\text{MS}}$ factorization scheme. Evolution equations for non-singlet quark DFFs are studied in Ref.~\cite{Majumder:2004wh}, while the singlet quark and gluon DFF evolution equations are studied in Ref.~\cite{Majumder:2004br}. In Refs.~\cite{Majumder:2004wh,Majumder:2004br} the ratio of the dihadron and single hadron fragmentation functions are examined, as this ratio is useful when considering experimental measurements. The choice of initial conditions is studied in Ref.~\cite{Grigoryan:2008ut}, primarily by considering the two-body correlation function.

Depending on the polarization of the fragmenting quark, special types of DFFs known as interference fragmentation functions (IFFs) can be constructed. The chiral-odd IFFs can be related to transversity~\cite{Boffi:1999it,Radici:1999sc,Radici:2002zg,Bacchetta:2004it,She:2007ht}. 
Transversity is one area of current interest that requires knowledge of the fragmentation functions of quarks~\cite{Barone:2001sp}. Out of the three leading-twist distribution functions that describe the quark structure of hadrons, it is the least well known, the other two being the unpolarized and helicity distributions. Recent work~\cite{Bacchetta:2011ip,Courtoy:2011ni,Courtoy:2012ry} suggests that DFFs may be useful in extracting transversity distributions by considering the SIDIS production of two hadrons with small invariant mass. Though transversity will not be the focus of this paper, it is presented as one possible motivation for further investigation into DFFs.

In this work we extend the latest version of the NJL-jet model, presented in Refs.~\cite{Matevosyan:2010hh,Matevosyan:2011ey}, to investigate dihadron fragmentation functions. In Se.~\ref{sec:qdff} we present a summary of fragmentation functions in the NJL-jet model, as set out in the aforementioned papers, with a focus on those parts that are relevant to understanding the dihadron fragmentation functions. Section~\ref{sec:DFF} outlines the extension of the NJL-jet model to be used in investigating DFFs, while results at the model scale for the DFFs are presented in Sec.~\ref{sec:results}. In Sec.~\ref{SEC_EVOL} we briefly discuss the QCD evolution equations for DFFs and present sample evolution results for our model.

\section{Quark Fragmentation Functions}
\label{sec:qdff}

This section provides a quick overview of the calculation of the quark fragmentation functions in the NJL-jet model~\cite{Ito:2009zc,Matevosyan:2010hh,Matevosyan:2011ey}, focusing on the aspects important to obtaining dihadron fragmentation functions within the model. Here we employ the $SU(3)$ NJL effective quark model~\cite{Nambu:1961tp,Nambu:1961fr,Kato:1993zw,Klimt:1989pm,Klevansky:1992qe} using light-cone (LC) coordinates~\cite{,Matevosyan:2010hh}. In the NJL model we include only the four-point quark interaction in the Lagrangian, with up, down, and strange quarks, and no additional free parameters. We employ Lepage-Brodsky (LB) ``invariant mass'' cutoff regularization for the loop integrals (see Refs.~\cite{Matevosyan:2010hh} for a detailed description as applied to the NJL-jet model).
\begin{figure}[H!b]
\begin{center}
\includegraphics[width=0.40\textwidth]{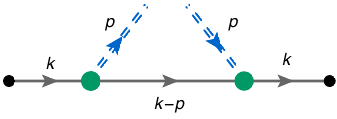}
\caption{Cut diagram for quark fragmentation function. Solid lines represent quarks and the double dashed lines a meson.}
\label{fig:quarkcut}
\end{center}
\end{figure}

The quark fragmentation function $D^h_q(z)$ is the probability for a quark of type $q$ to emit a hadron of type $h$ carrying fraction $z$ of its light-cone momentum $h$ (here meson $m=q\bar{Q}$). We denote the elementary quark fragmentation function, corresponding to the situation where the detected hadron is the only emitted hadron, by $d^h_q(z)$. The corresponding cut diagram for the elementary quark fragmentation function is shown in Fig.~\ref{fig:quarkcut}. The elementary fragmentation function depicted in Fig.~\ref{fig:quarkcut} can be written as
\begin{widetext}
\begin{eqnarray}
\label{eqff}
d^m_q(z) & = & N_c\frac{C^m_q}{2}g^2_{mqQ}\frac{z}{2}\int\frac{d^4k}{(2\pi)^4}Tr[S_1(k)\gamma_{+}S_1(k)\gamma_5(k\!\!\!/-p\!\!\!/+M_2)\gamma_5]\times \delta(k_--p_-/z)2\pi\delta((p-k)^2-M^2_2) \nonumber \\
& = & \frac{C^m_q}{2}g^2_{mqQ}z\int\frac{d^2p_\perp}{(2\pi)^3}\frac{p_\perp^2+((z-1)M_1+M_2)^2}{(p_\perp^2+z(z-1)M_1^2+zM_2^2+(1-z)m_m^2)^2},
\end{eqnarray}
\end{widetext}
where $C^m_q$ is the corresponding flavor factor and $g_{mqQ}$ is the quark-meson coupling. The masses $M_1$, $M_2$ and $m_m$ are the masses of the fragmenting quark, the remnant quark and the produced hadron (here the hadron is a meson), respectively.

If a sharp cutoff in the transverse momentum, $P_\perp^2$, is assumed, the integration in Eq.~(\ref{eqff}) can be evaluated analytically~[Eq.~\eqref{aeff}].
\begin{equation}
\label{aeff}
d^m_q(z) = \frac{C^m_q}{2}\frac{g^2_{mqQ}}{8\pi^2}z\left(\frac{A/B-1}{B/P^2_\perp+1}+\ln(1+P^2_\perp/B)\right),
\end{equation}
where
\begin{eqnarray}
A & \equiv & ((z-1)M_1+M_2)^2,\\
B & \equiv & z(z-1)M_1^2+zM_2^2+(1-z)m^2_m.
\end{eqnarray}

The Lepage-Brodsky ``invariant mass" cutoff regularization method (Refs.~\cite{Ito:2009zc} and~\cite{Bentz:1999gx} describe this when applied to the NJL-Jet model) is employed to regularize the loop integrals. The loop integrals are regularized by setting a cutoff on the invariant mass, $M_{12}$, such that
\begin{equation}
M_{12}\le\Lambda_{12}\equiv\sqrt{\Lambda^2_3+m^2_m}+\sqrt{\Lambda^2_3+M^2_2},
\end{equation}
where $\Lambda_{12}$ is the maximum invariant mass. Here the $3$-momentum cutoff, denoted by $\Lambda_3$, is fixed by reproducing the value of the experimentally measured pion decay constant.  In Lepage-Brodsky regularization, $P_\perp^2$ is given by:
\begin{eqnarray}
P^2_\perp & = & z(1-z)\left(\sqrt{\Lambda^2_3+m^2_m}+\sqrt{\Lambda^2_3+M^2_2}\right)\nonumber\\
& & -(1-z)m^2_m-zM^2_2.
\end{eqnarray}

The value of the 3-momentum cutoff used in this work, $\Lambda_3=0.67 \text{ GeV}$, was obtained in Ref.~\cite{Bentz:1999gx} by choosing the constituent light quark mass $M=0.3 \text{ GeV}$ and using pion decay $f_\pi=0.093$. The corresponding  constituent strange quark mass, $M_s=0.537 \text{ GeV}$, was obtained by reproducing the pion and kaon masses, $m_\pi=0.14\text{ GeV}$ and $m_K=0.495\text{ GeV}$. The calculated values of the quark-meson couplings are $g_{\pi qQ}=3.15$ and $g_{KqQ}=3.3876$.

The elementary quark fragmentation function, integrated over the light-cone momentum fraction $z$, represents the total probability of a quark splitting into a hadron of a given type plus another quark. In construction of the NJL-jet model we are interested in procesess where a hadron is produced at each step. Thus we construct the renormalized elementary fragmentation functions, $\hat{d}^m_q(z)$, such that the total probability of emitting a hadron (summed over all possible hadron types $m^\prime$ that the quark $q$ can emit in the elementary splitting process) is one:
\begin{equation}
\hat{d}^m_q(z)=\frac{d^m_q(z)}{\displaystyle\sum_{m^\prime}\int^1_0 d^{m^\prime}_q(z)}.
\end{equation}

\begin{figure}[H!t]
\begin{center}
\includegraphics[width=0.45\textwidth]{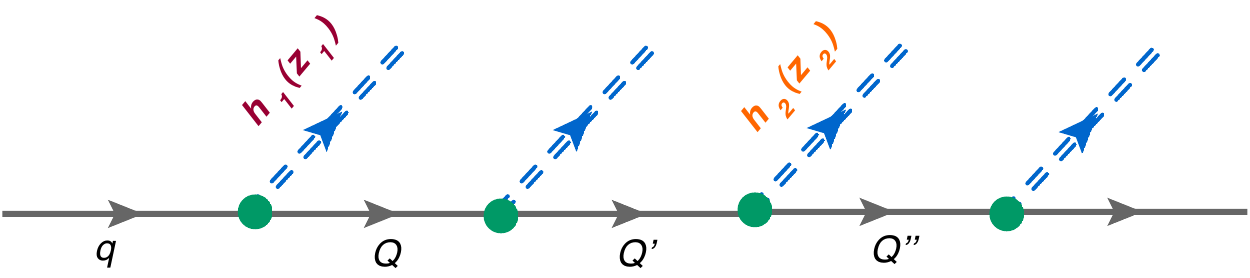}
\caption{Quark cascade} \label{fig:quarkcascade}
\end{center}
\end{figure}

In a quark-jet-model, the total fragmentation function, $D^m_q(z)$, is described by successive elementary splittings of a quark into hadrons. This process is shown diagrammatically in Fig.~\ref{fig:quarkcascade}. The initial quark $q$ fragments into a meson, $m=q\bar{Q}$, with light-cone momentum fraction $z$ of the initial quark's momentum, and a quark, $Q$, with light-cone momentum fraction $1-z$. The emitted quark $Q$ fragments as well, and the process repeats, forming a cascade of hadrons. It is important to note that within the model the emitted hadrons do not interact with the other hadrons produced in the quark jet. An integral equation for the quark cascade process shown in Fig.~\ref{fig:quarkcascade} was derived in the quark-jet model of Ref.~\cite{Field:1977fa}. The integral equation for the total fragmentation function is
\begin{equation}
D^m_q(z)  =  \hat{d}^m_q(z)+\sum_{Q}\int^{1}_{z}\frac{dy}{y}\hat{d}^Q_q(\frac{z}{y})D^m_Q(y),
\label{EQ_SINGEL_FRAG}
\end{equation}
where $\hat{d}^Q_q(z)  =  \hat{d}^m_q(1-z)|_{m=q\bar{Q}}$.

The probabilistic interpretation of Eq.~(\ref{EQ_SINGEL_FRAG}) can be clarified by multiplying both sides by a factor of $dz$. The term on the left-hand side is the probability for the quark $q$ to emit meson $m$ with light-cone momentum fraction $z$. On the right-hand side, the first term is the driving function, which represents the probability of creating a meson $m$ carrying momentum fraction $z$ to $z+dz$ from the first emission step and the second term represents the probability of creating the meson, $m$, further in the quark decay chain. The above equation is solved by uniformly discretizing $z$ and $y$ in the interval $[0,1]$ and approximating the integrals as sums over these discrete values of $z$ and $y$. Then $D^m_q(z)$ and $\hat{d}^m_q(z)$ can be expressed as vectors $\vec{D}^m_q$ and $\vec{f}$ of values at the discretization points of $z$, and the integrand of the second term,  without $D^m_q(y)$,  can be written as a matrix $g$ over the values of the discretization points of $z$ and $y$:
\begin{eqnarray}
&\vec{D}^m_q =  \vec{f}+g\cdot\vec{D}^m_q &\nonumber\\
&(I-g)\cdot\vec{D}^m_q =  \vec{f}&\nonumber\\
&\vec{D}^m_q =  (I-g)^{-1}\vec{f}&,
\label{deriveD}
\end{eqnarray}
 where $I$ is the unit matrix.

\begin{figure}[H!b]
\begin{center}
\includegraphics[width=0.5\textwidth]{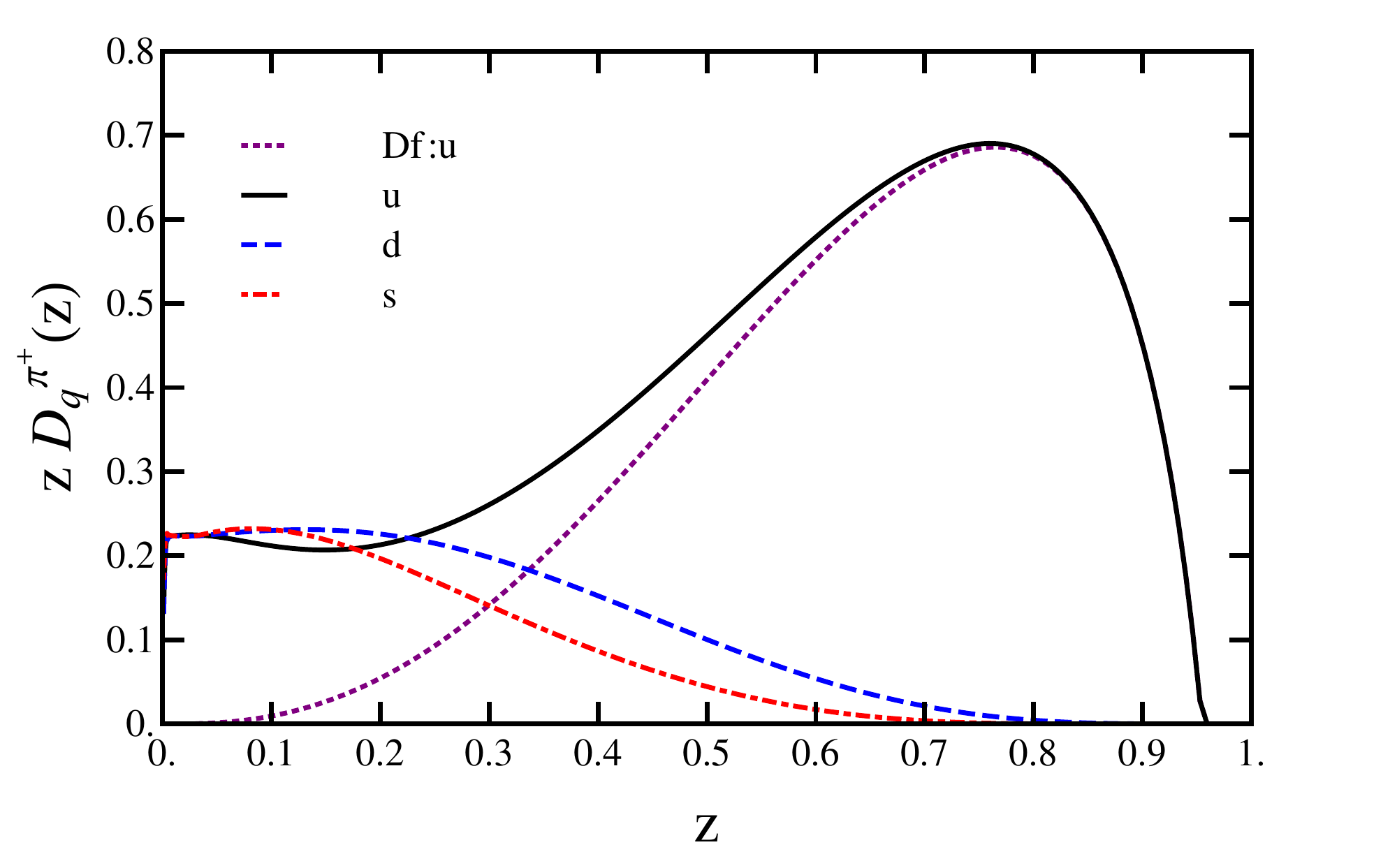}
\caption{Fragmentation functions of $u$~(black solid line), $d$~(blue dashed line) and $s$~(red dot-dashed line) quarks to $\pi^+$ as a function of the light-cone momentum fraction $z$. Driving function of the $u$~(purple dotted line) to $\pi^+$ is labeled as Df:$u$.} \label{fig:SFFresults}
\end{center}
\end{figure}

Here it is important to use an appropriate number of discretization points to avoid large numerical errors when solving for $D_q^m(z)$. The number of points used was increased until there was sufficient convergence of the solutions of the fragmentation functions. The resulting solutions for the fragmentation functions of $u$, $d$ and $s$ quarks to $\pi^+$ and $K^+$ are presented in Ref.~\cite{Matevosyan:2011ey}. The fragmentation functions of $u$, $d$ and $s$ quarks to $\pi^+$ are also shown here in Fig.~\ref{fig:SFFresults}, as well as the driving function of the $u$ to $\pi^+$, $\hat{d}^{\pi^+}_u$, which is labeled as Df:$\,u$ (the notation Df:$\,q$ will be adopted in Sec.~\ref{sec:dfcontri} as well).

\section{Dihadron Fragmentation Functions}
\label{sec:DFF}
We now consider a semi-inclusive process in which two hadrons are detected in the final state. This requires a new fragmentation function, known as the dihadron fragmentation function, that describes the probability of this process. DFFs may be useful in the extraction of transversity distributions~\cite{Bacchetta:2011ip}, which are the least well known of the three leading-twist distribution functions that describe the quark structure of hadrons, the other two being the unpolarized and helicity distribution functions. We now extend the NJL-jet model to describe the DFFs. Dihadron fragmentation functions, $D^{h_1,h_2}_q(z_1,z_2)$, correspond to the probability of a quark $q$ producing two hadrons, $h_1$ and $h_2$, that carry its light-cone momentum fractions $z_1$ and $z_2$, respectively. An illustration of how a quark cascade can produce two observed hadrons, $h_1$ and $h_2$, in the NJL-jet model is shown in Fig.~\ref{fig:quarkcascade}.

The integral equation for the dihadron fragmentation function $D^{h_1,h_2}_q(z_1,z_2)$ has been constructed by Field and Feynman in the quark-jet model [Eqs (2.43a)-(2.43d) of Ref.~\cite{Field:1977fa}], which is shown in Eq.~(\ref{DFF2}). Here $\hat{d}^h_q(z)$ and $\hat{d}^Q_q(\eta)$ are the elementary splitting functions of the quark $q$ to the corresponding hadron $h$ and quark $Q$. On the left-hand side of Eq.~(\ref{DFF2}) is the term representing the probability for the quark $q$ to emit hadrons $h_1$ and $h_2$ with light-cone momentum fractions $z_1$ and $z_2$, respectively. The first term on the right-hand side of Eq.~(\ref{DFF2}) corresponds to the probability of producing hadron $h_1$ from the quark $q$ at the first step in the cascade, followed by hadron $h_2$ produced either directly afterwards or further down in the quark decay chain, while the second term is similar to the first one, except for $h_1 \leftrightarrow h_2$ . The third term corresponds to the probability of having both $h_1$ and $h_2$ produced after the first hadron emission.
\begin{widetext}
\begin{equation}
\label{DFF2}
D^{h_1,h_2}_q(z_1,z_2)=\hat{d}^{h_1}_q(z_1)\frac{D^{h_2}_{q_1}(\frac{z_2}{1-z_1})}{1-z_1}+\hat{d}^{h_2}_q(z_2)\frac{D^{h_1}_{q_2}(\frac{z_1}{1-z_2})}{1-z_2}+\sum_Q\int^1_{z_1+z_2}\frac{d\eta}{\eta^2}\hat{d}^Q_q(\eta)D^{h_1,h_2}_Q\left(\frac{z_1}{\eta},\frac{z_2}{\eta}\right),
\end{equation}
\begin{equation}
q\rightarrow h_1+q_1;\quad q\rightarrow h_2+q_2.
\end{equation}
\end{widetext}

 In the integral term we perform a change of integration variables to $\xi_1=z_1/\eta$ and $\xi_2=z_2/\eta$, so that the arguments of $D^{h_1,h_2}_Q(\xi_1,\xi_2)$ will correspond to $\xi_1$ and $\xi_2$ at grid point values when uniformly discretized:
\begin{widetext}
\begin{eqnarray}
& & \int^1_{z_1+z_2}\frac{d\eta}{\eta^2}\hat{d}^Q_q(\eta)D^{h_1,h_2}_Q\left(\frac{z_1}{\eta},\frac{z_2}{\eta}\right)\nonumber \\ 
&  & =\int^{\frac{z_1}{z_1+z_2}}_{z_1}d\xi_1\int^{\frac{z_2}{z_1+z_2}}_{z_2}d\xi_2\int^1_{z_1+z_2}d\eta\frac{\delta(\xi_1-z_1/\eta)}{\eta}\frac{\delta(\xi_2-z_2/\eta)}{\eta}\hat{d}^Q_q(\eta)D^{h_1,h_2}_Q(\xi_1,\xi_2)\nonumber\\ 
&  & =\int^{\frac{z_1}{z_1+z_2}}_{z_1}d\xi_1\int^{\frac{z_2}{z_1+z_2}}_{z_2}d\xi_2\int^1_{z_1+z_2}d\eta\delta(z_1-\xi_1\eta)\delta(z_2-\xi_2\eta)\hat{d}^Q_q(\eta)D^{h_1,h_2}_Q(\xi_1,\xi_2) \nonumber\\
&  & =\int^{\frac{z_1}{z_1+z_2}}_{z_1}d\xi_1\int^{\frac{z_2}{z_1+z_2}}_{z_2}d\xi_2\delta(z_2\xi_1-z_1\xi_2)\hat{d}^Q_q(z_1/\xi_1)D^{h_1,h_2}_Q(\xi_1,\xi_2) 
\end{eqnarray}
Then the equation for the dihadron fragmentation functions takes the following form:
\begin{eqnarray}
\label{DFFcv}
D^{h_1,h_2}_q(z_1,z_2) & = & \hat{d}^{h_1}_q(z_1)\frac{D^{h_2}_{q_1}(\frac{z_2}{1-z_1})}{1-z_1}+\hat{d}^{h_2}_q(z_2)\frac{D^{h_1}_{q_2}(\frac{z_1}{1-z_2})}{1-z_2}\nonumber\\
& & +\sum_Q\int^{\frac{z_1}{z_1+z_2}}_{z_1}d\xi_1\int^{\frac{z_2}{z_1+z_2}}_{z_2}d\xi_2\delta(z_2\xi_1-z_1\xi_2)\hat{d}^Q_q(z_1/\xi_1)D^{h_1,h_2}_Q(\xi_1,\xi_2).
\end{eqnarray}
\end{widetext}

To solve the above equation for the dihadron fragmentation function $D^{h_1,h_2}_q(z_1,z_2)$, we discretize $z_1$, $z_2$, $\xi_1$, and $\xi_2$ uniformly in the interval $[0,1]$ and approximate the integrals as sums over the discretized values of these variables. The fragmentation functions are written in matrix form, where the elements of the matrices are their values at the corresponding uniformly discrete values of the arguments. We used Mathematica to solve for both the single hadron and dihadron fragmentation functions. The number of discretization points used for the single hadron fragmentation functions was $500$, while the number of the discretization points afforded for the dihadron fragmentation functions was $200$. These values for the numbers of discretization points produced convergence of the solutions within typically $5\%$, while allowing for a reasonable computational time and computer memory size required by the problem. Several techniques were used to lower the memory use of the program, including the use of the sparse arrays in Mathematica. To calculate the third term of Eq.~(\ref{DFFcv}), the integrals over $\xi_1$ were converted to a sum over its uniformly discrete values. The delta function was used to eliminate the integration over $\xi_2$. The values of $\xi_2$ that are selected by the delta function may not match any of its uniformly discretized values. To account for this, the values of the DFFs at the selected $\xi_2$ were obtained using linear interpolation from neighboring discrete values.

The advantage of the approach presented here is that there is a single underlying effective quark model description at the microscopic level for both parton distribution functions and fragmentation functions, with no fitted parameters to fragmentation data. Moreover, recent developments of the model for the single quark fragmentations allow us to extend the model using Monte Carlo techniques to describe the production of hadronic resonances and the inclusion of transverse momenta~\cite{Matevosyan:2011ey,Matevosyan:2011vj}. In the future, this and other extensions of the model can also be incorporated for the DFFs.

\section{Results}
\label{sec:results}
In this section we investigate various features of the DFFs obtained as solutions of Eq.~(\ref{DFFcv}). Section~\ref{sec:dfcontri} investigates the contribution of the integral term versus that of the driving term to the solution of DFF. The impact of the inclusion of the strange quark on $D^{\pi^+\pi^-}_{u}$ and $D^{\pi^+\pi^-}_{u}$ is studied in Sec.~\ref{sec:strangecontri}. In Sec.~\ref{sec:zfixed}, we consider $D^{\pi^+K^-}_{q}(z_1,z_2)$ and fix either $z_1$ or $z_2$, to study the dependence of this DFF on each of the variables.

\subsection{Contribution of the integral term}
\label{sec:dfcontri}

In Eq.~(\ref{DFFcv}), the sum of the first two terms is considered to be the driving function of the dihadron fragmentation function and they describe the probability of emitting one of the detected hadrons in the first emission step. The last term in Eq.~(\ref{DFFcv}) corresponds to the probability of emitting both detected hadrons after the emission of a hadron in the first step. We now consider the contribution of this last term to the solution for DFF by comparing them with the corresponding driving functions for three combinations of observed pions and kaons: $\pi^+\pi^-$, $\pi^+K^-$ and $K^+K^-$.

\begin{figure}[H!t]
\begin{center}
\subfigure[$z_1=0.5$: $z_2\,D^{\pi^+\pi^-}_q$]{\label{fig:df_uds_pplpmi_z1eq0_5}
\includegraphics[width=0.48\textwidth]{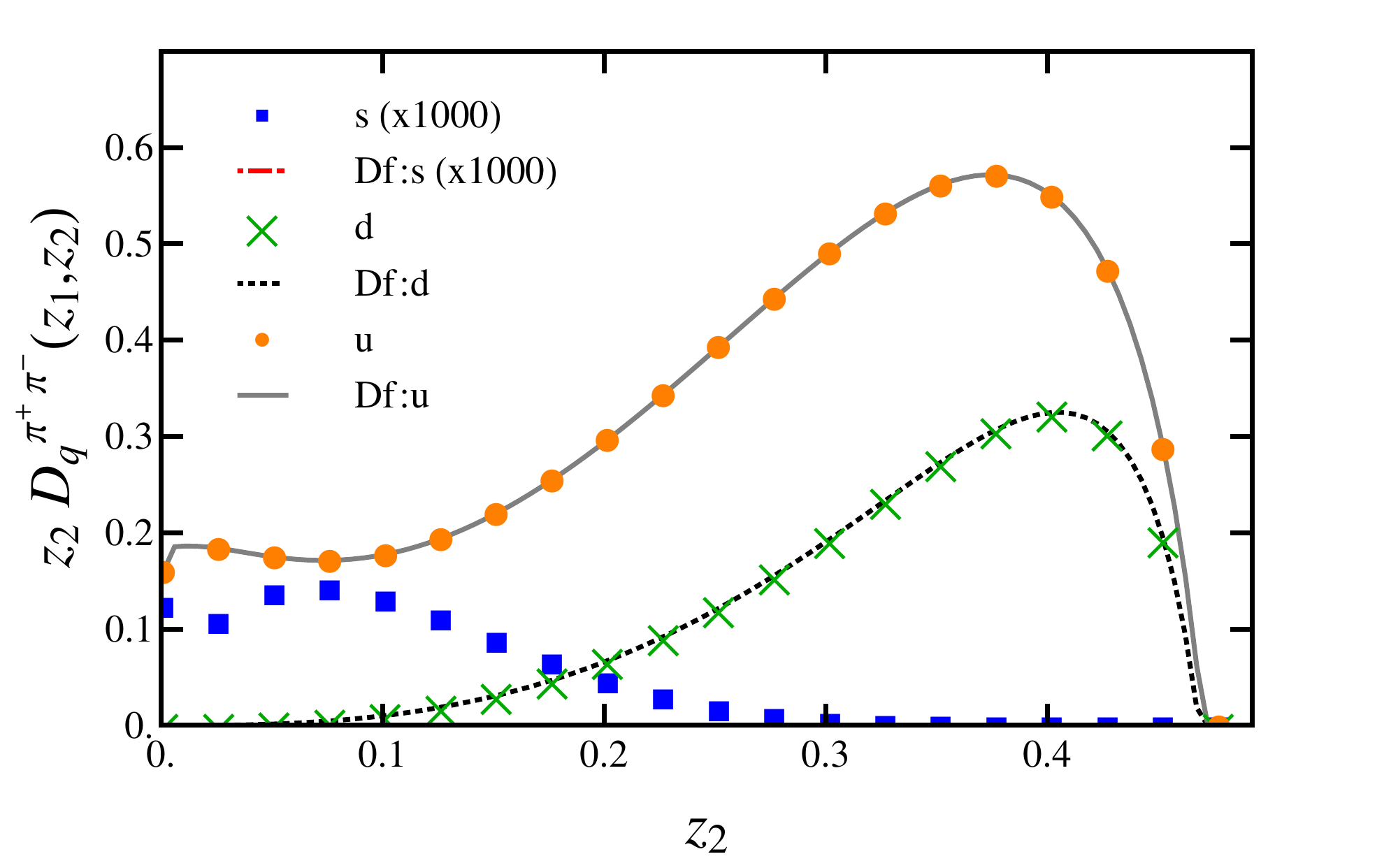}}
\subfigure[$z_1=0.5$: $z_2\,D^{\pi^+K^-}_q$]{\label{fig:df_uds_pplkmi_z1eq0_5}
\includegraphics[width=0.48\textwidth]{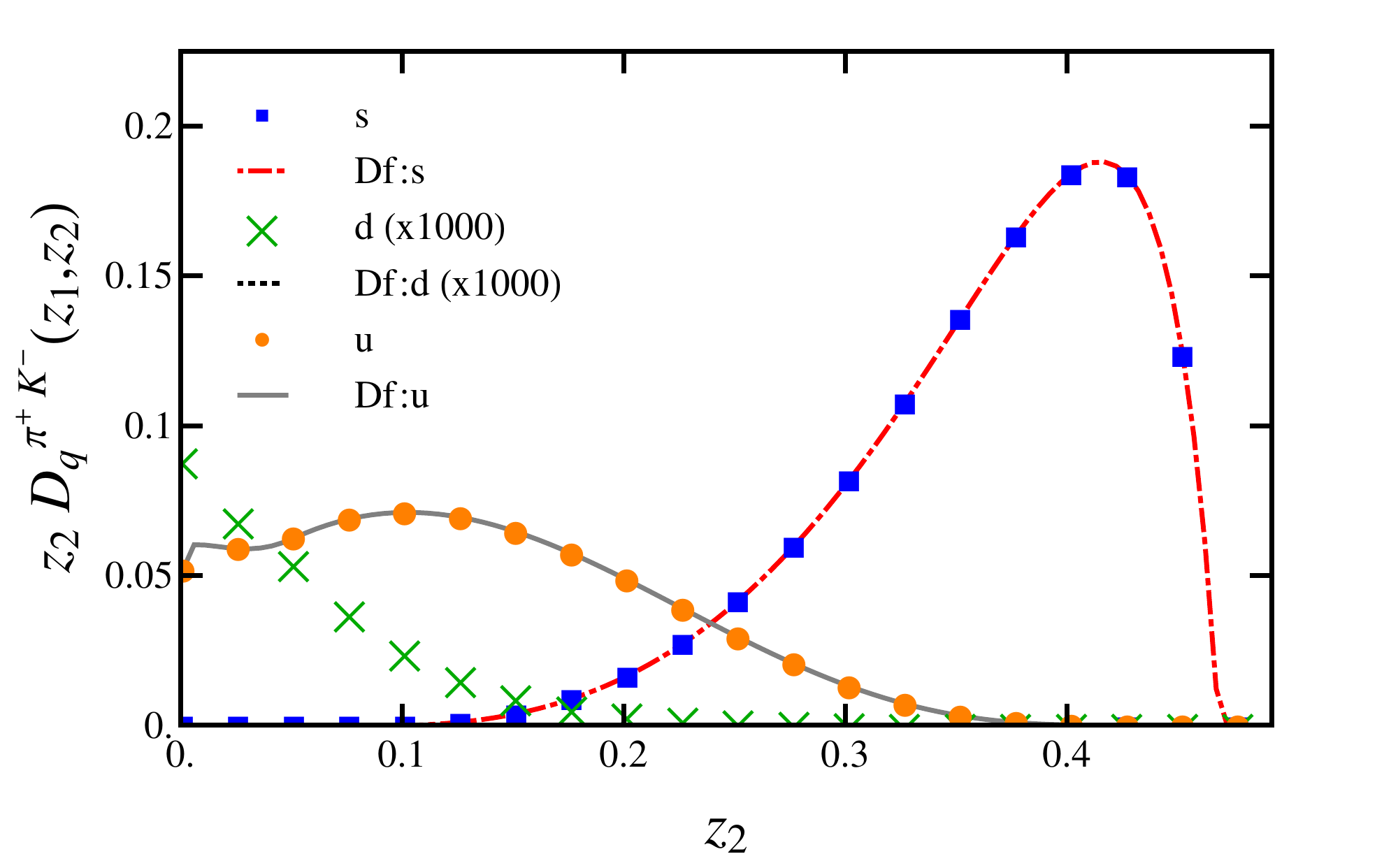}}       
\subfigure[$z_1=0.5$: $z_2\,D^{K^+K^-}_q$]{\label{fig:df_uds_kplkmi_z1eq0_5}
\includegraphics[width=0.48\textwidth]{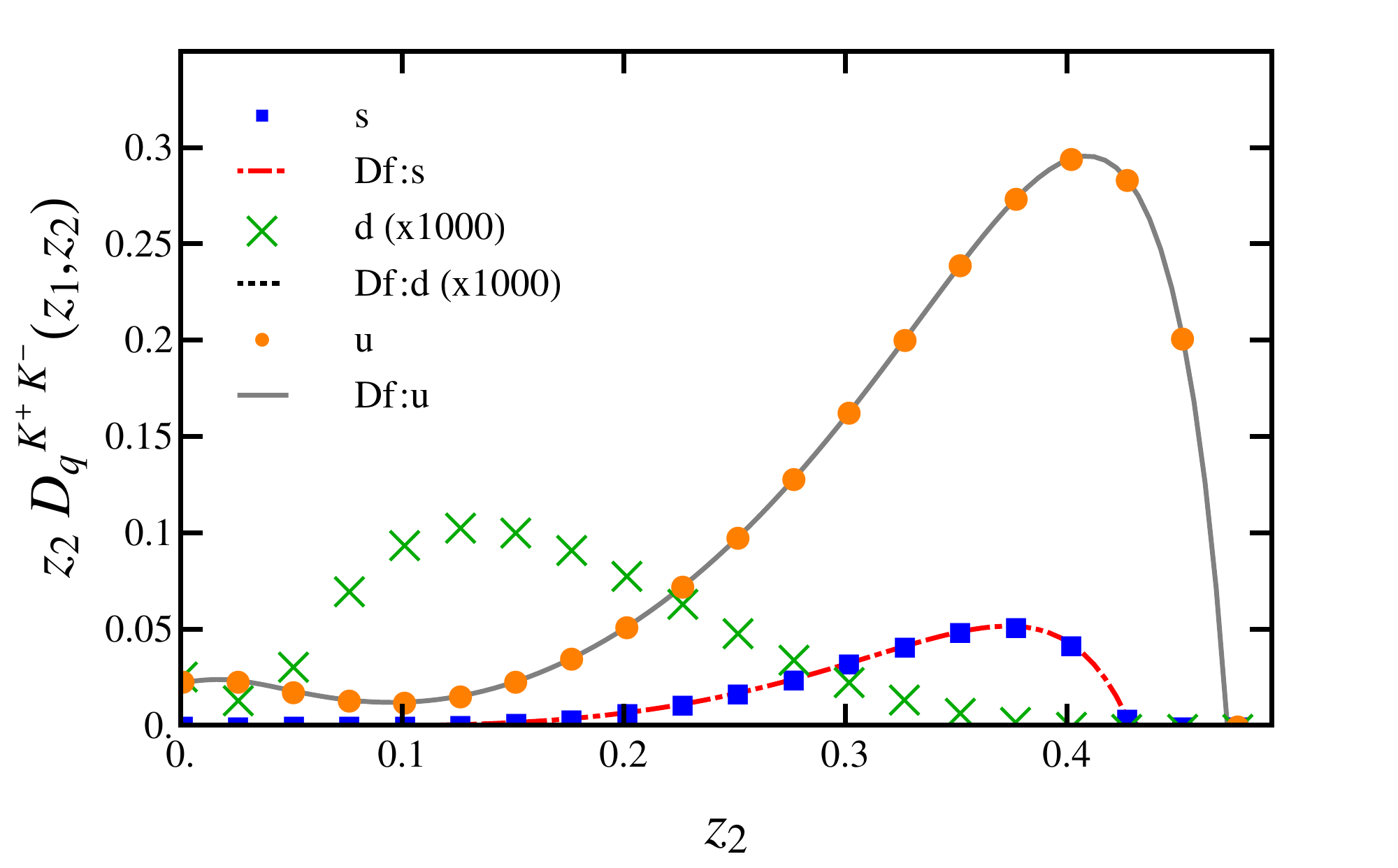}}       
\caption{Dihadron fragmentation functions for $z_1=0.5$ for~\subref{fig:df_uds_pplpmi_z1eq0_5} $h_1=\pi^+, h_2=\pi^-$,~\subref{fig:df_uds_pplkmi_z1eq0_5} $h_1=\pi^+, h_2=K^-$ and~\subref{fig:df_uds_kplkmi_z1eq0_5} $h_1=K^+, h_2=K^-$. The DFFs and driving functions of the up, down and strange quarks are shown by the orange circle points and solid gray line, green diamond points and dotted black line, and the blue square points and the dot-dashed red line, respectively. Driving functions for fragmenting quark $q$ are also labeled in the legend as Df:$q$. The number in the brackets in the legend indicates the scaling factor used in depicting the curve.}
\label{fig:drivfunc05}
\end{center}
\end{figure}

The DFFs for the produced hadrons ${\pi^+\pi^-}$, ${\pi^+K^-}$ and ${K^+K^-}$ as functions of $z_2$ with $z_1=0.5$ are shown in Fig.~\ref{fig:drivfunc05}. The plots in the figures show that the favored DFFs, where the initial quark can produce either of the detected hadrons from the initial quark, are almost equal to the driving function, with the integral term giving only a very small contribution. The unfavored DFFs, where neither of the hadrons can be directly produced by the initial quark, are generated entirely by the integral term.

\begin{figure}[H!t]
\begin{center}
\subfigure[$z_1=0.1$: $z_2\,D^{\pi^+\pi^-}_q$]{\label{fig:df_uds_pplpmi_z1eq0_1}
\includegraphics[width=0.48\textwidth]{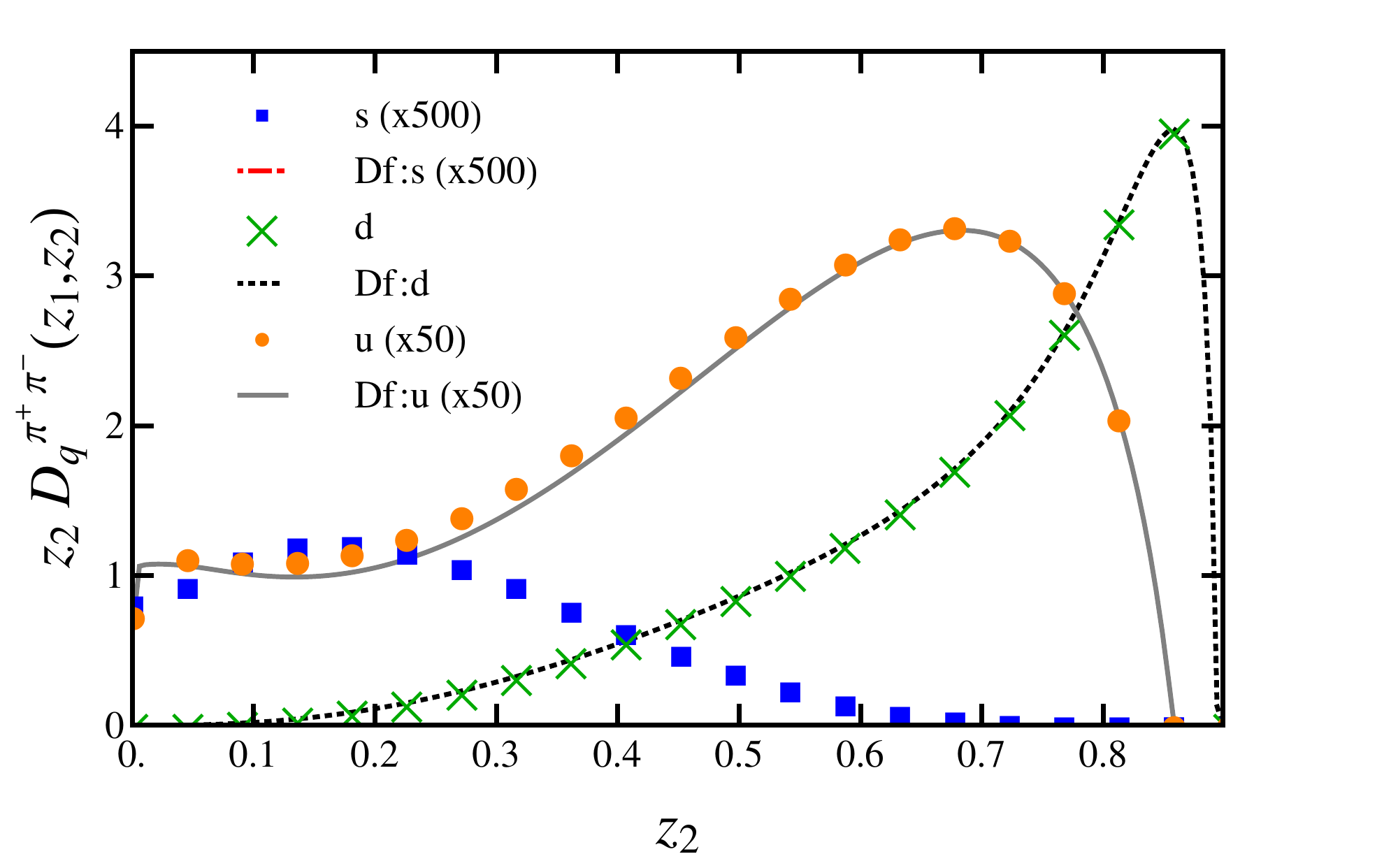}}
\subfigure[$z_1=0.1$: $z_2\,D^{\pi^+K^-}_q$]{\label{fig:df_uds_pplkmi_z1eq0_1}
\includegraphics[width=0.48\textwidth]{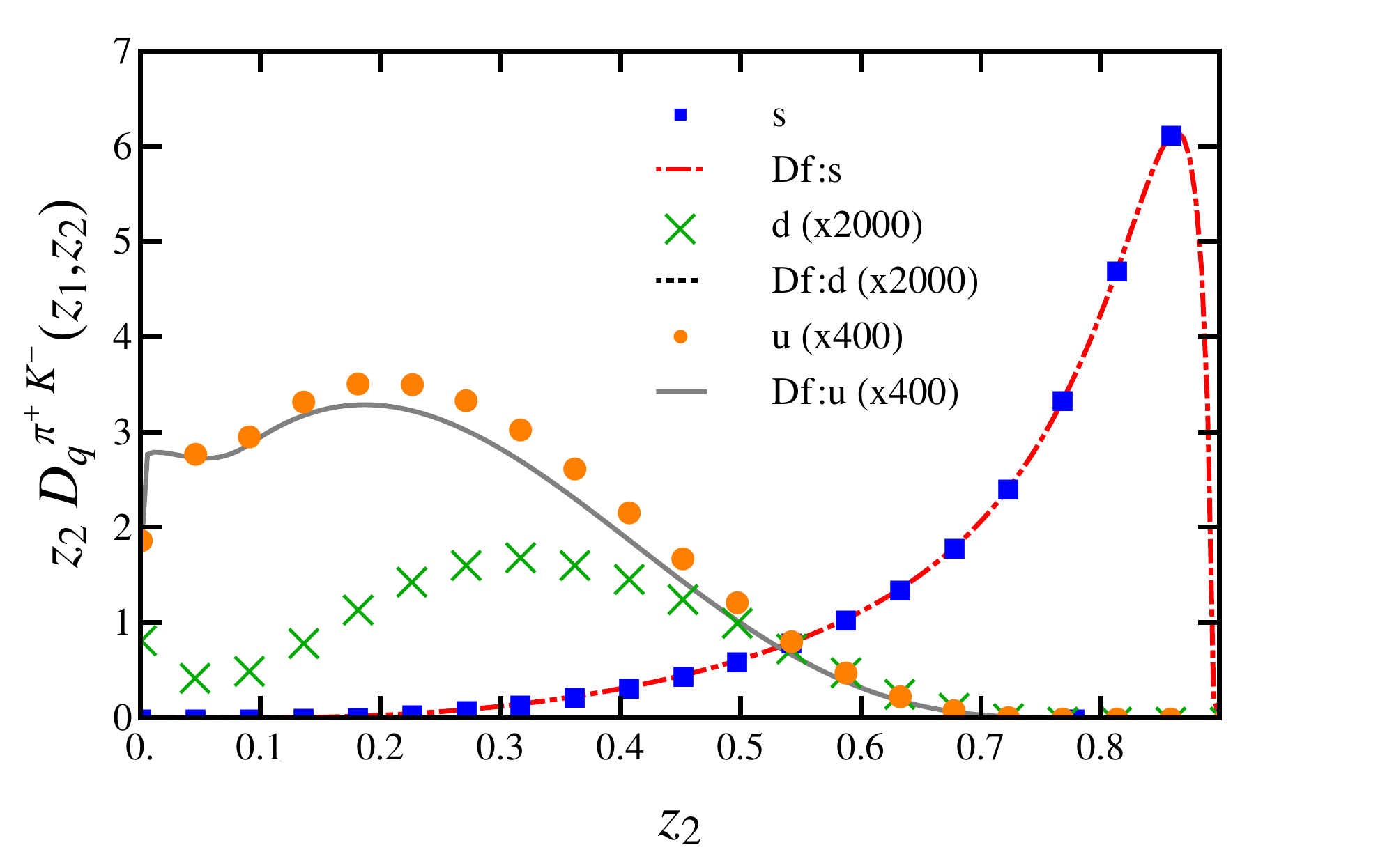}}       
\subfigure[$z_1=0.1$: $z_2\,D^{K^+K^-}_q$]{\label{fig:df_uds_kplkmi_z1eq0_1}
\includegraphics[width=0.48\textwidth]{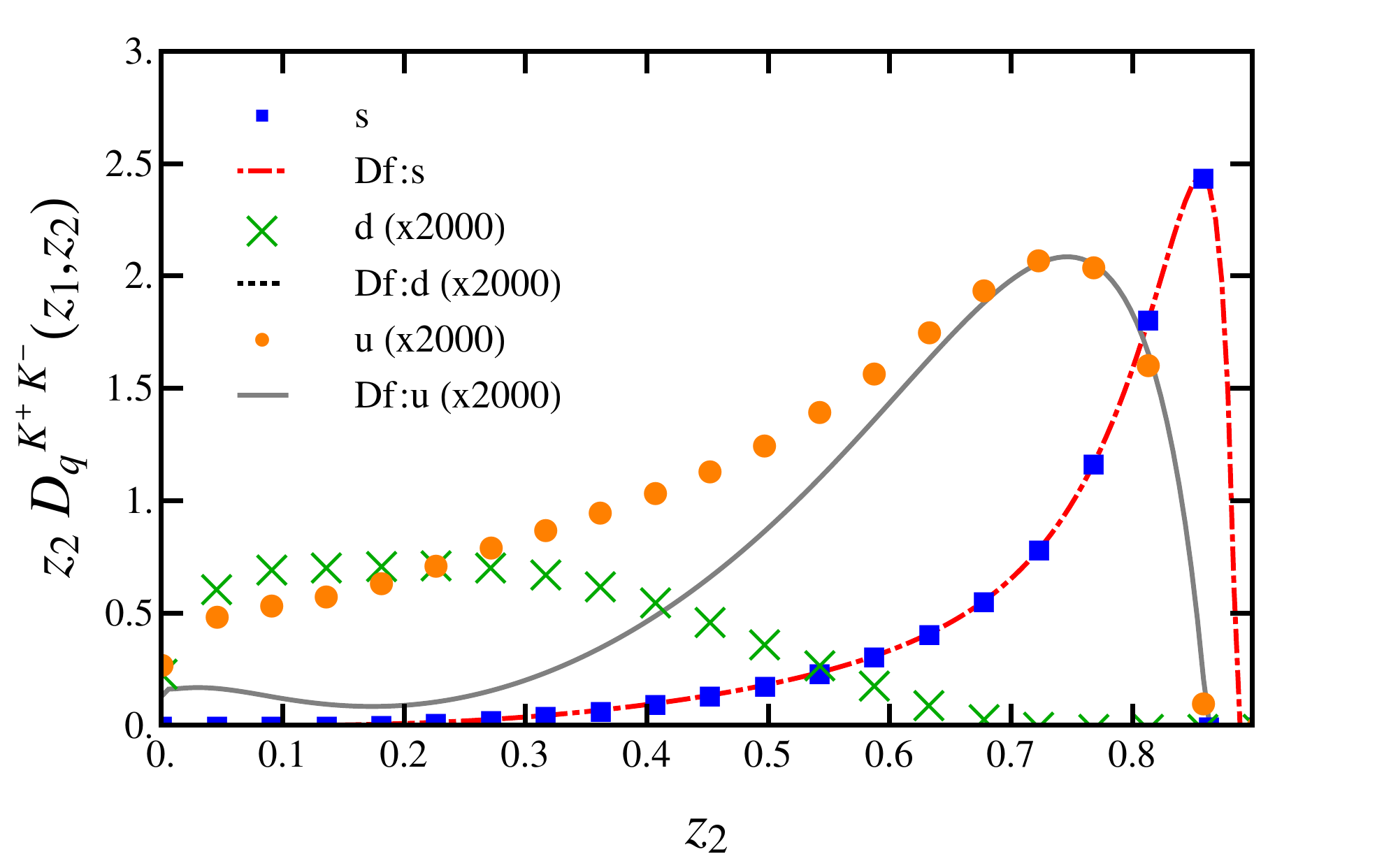}}
\caption{Dihadron fragmentation functions for $z_1=0.1$ for~\subref{fig:df_uds_pplpmi_z1eq0_1} $h_1=\pi^+, h_2=\pi^-$,~\subref{fig:df_uds_pplkmi_z1eq0_1} $h_1=\pi^+, h_2=K^-$ and~\subref{fig:df_uds_kplkmi_z1eq0_1} $h_1=K^+, h_2=K^-$. The DFFs and driving functions of the up, down and strange quarks are shown by the orange circle points and solid gray line, green diamond points and dotted black line, and the blue square points and and the dot-dashed red line, respectively. Driving functions for fragmenting quark $q$ are also labeled in legend as Df:$q$. The number in the brackets in the legend indicates the scaling factor used in depicting the curve.}
\label{fig:drivfunc01}
\end{center}
\end{figure}

In Figs.~\ref{fig:drivfunc05} and~\ref{fig:drivfunc01}, the solution of the DFF for the up quark is shown by the orange circle points and the driving function is shown as a solid gray line. The green diamond points and dotted black line show the DFF and the driving function for the down quark, respectively. The DFF and the driving function for the strange quark are shown by the blue square points and the dot-dashed red line, respectively. In each of the figures, the number in the brackets in the corresponding legend indicate the scaling factor used in depicting the curve on the plot. This notation for the scaling factor is also used in Sec.~\ref{sec:strangecontri} and~\ref{sec:zfixed}.

In Fig.~\ref{fig:drivfunc01}, the results for the DFFs and driving functions for the $\pi^+\pi^-$, $\pi^+K^-$ and $K^+K^-$ are presented for the fixed value of $z_1=0.1$. Here the integral term contribution to the up quark DFFs become visible as the value of $z_1$ is lowered because the driving function's contribution to the DFF becomes less significant. The driving function's contribution for the down~($\to\pi^+\pi^-$) and strange quark~($\to\pi^+K^-$ and $K^+K^-$) DFFs are still very dominant, so there is no noticeable contribution from the integral term here. It is worth noting that the integral term contributions to both favored and unfavored DFFs are of the same magnitude, but the contributions to the favored DFFs are only noticeable when the driving function is not dominating the solution.  We note also that the integral equations of $D^{\pi^+ \pi^-}_u$ and $D^{\pi^+ \pi^-}_d$ are symmetric in $z_1 \leftrightarrow 
z_2$, such that the integral equation of $q=u$ for fixed $z_1$ equals the integral equation of $q=d$ for fixed $z_2$~($D^
{\pi^+ \pi^-}_u(z_1,z_2)=D^{\pi^+ \pi^-}_d(z_2,z_1)$). In Sec.~IV C, we will use the same fixed values for $z_1$ and $z_2$ when examining $D^{\pi^+ K^-}_q$, as this flavor symmetry is absent there.

\subsection{Impact of including the strange quark on the $D^{\pi^+\pi^-}_q$}
\label{sec:strangecontri}
\begin{figure}[H!t]
\begin{center}
\subfigure[$z_1=0.1$: $z_2\,D^{\pi^+\pi^-}_q$]{
\label{fig:snos_ud_pplpmi_z1eq0_1}
\includegraphics[width=0.48\textwidth]{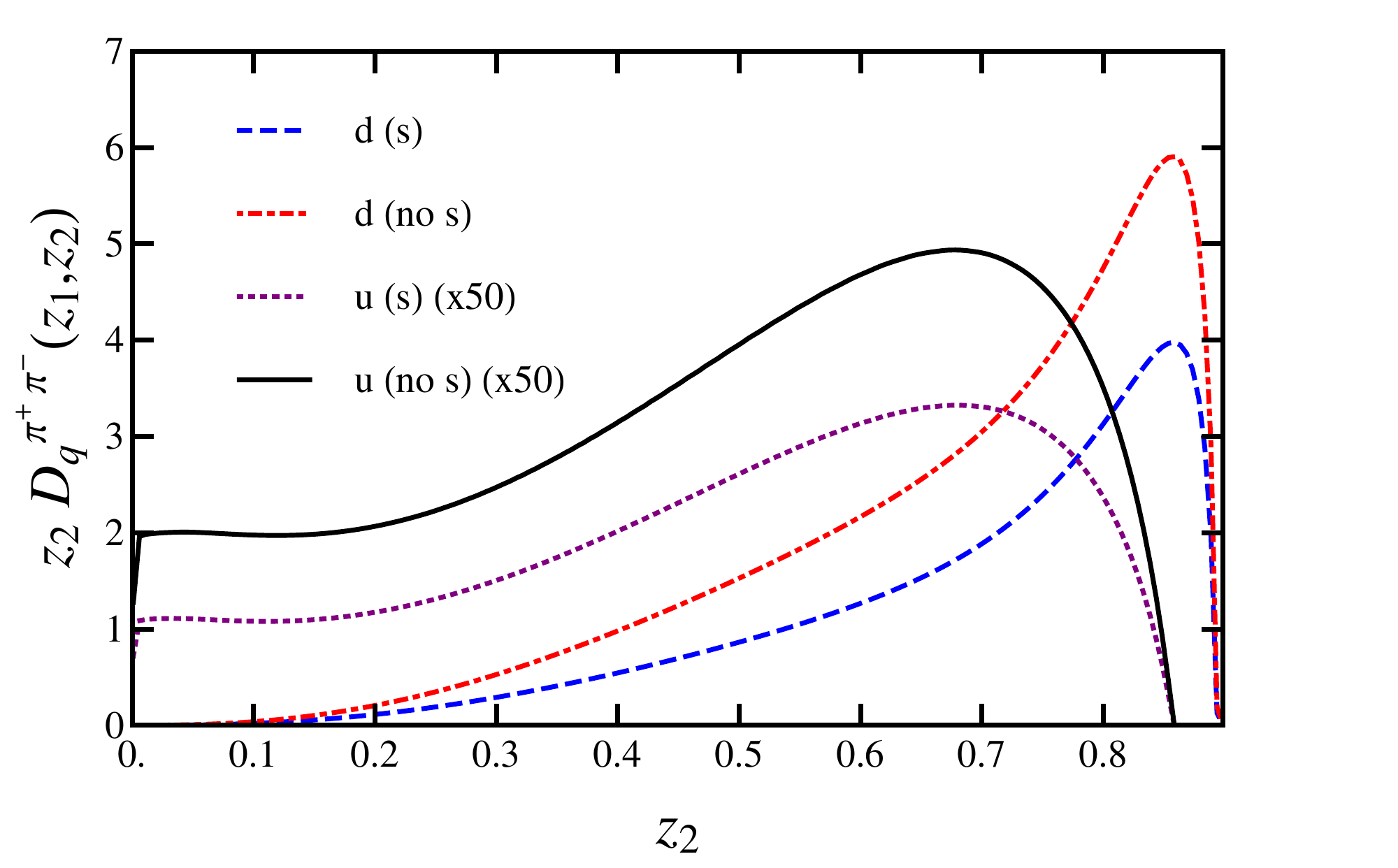}}            
\subfigure[$z_1=0.5$: $z_2\,D^{\pi^+\pi^-}_q$]{
\label{fig:snos_ud_pplpmi_z1eq0_5}
\includegraphics[width=0.48\textwidth]{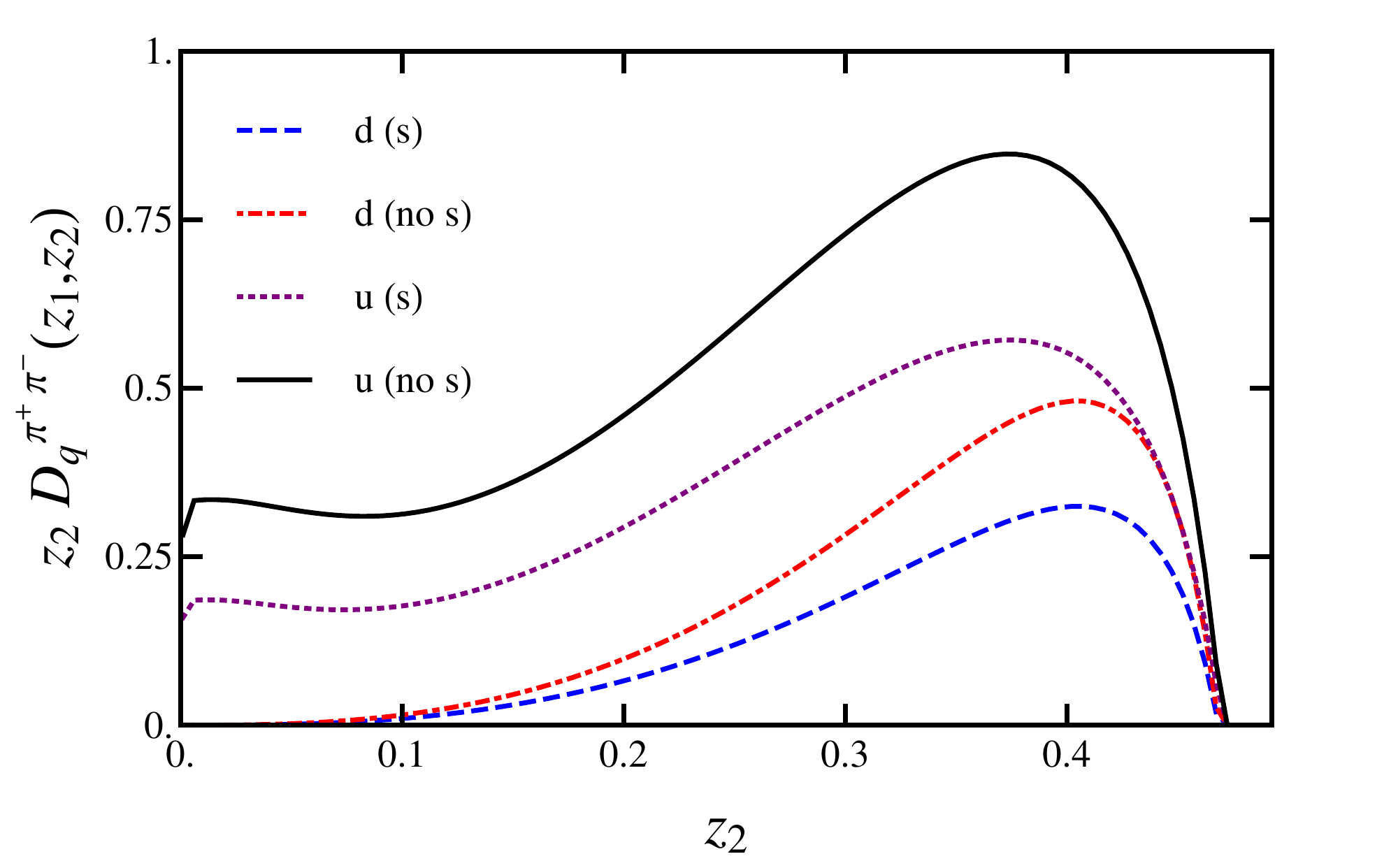}}
\caption{Comparison of strange quark contribution of $\pi^+\pi^-$ dihadron fragmentation functions for~\subref{fig:snos_ud_pplpmi_z1eq0_1} $z_1=0.1$ and~\subref{fig:snos_ud_pplpmi_z1eq0_5} $z_1=0.5$. The dashed blue line and the red dot-dashed line represent the results for up quark DFF with and without the strange quark~[denoted by (s) and (no s) in the captions], respectively. Similarly, the purple dotted line and the black solid line represent the results for the down quark DFF with and without the strange quark, respectively. The number in the brackets in the legend indicates the scaling factor used in depicting the curve.}
\label{fig:dffsquark}
\end{center}
\end{figure}
We now study the impact of the inclusion of the strange quark on $D^{\pi^+\pi^-}_{u}$ and $D^{\pi^+\pi^-}_{d}$. In Eq.~(\ref{DFFcv}) the integral term contains a sum over $Q$ that runs over the flavors of the quarks considered in the model, thus the inclusion of the strange quark couples the DFFs for the $u$ and $d$ quarks to those of the $s$ quark. Also, the inclusion of the strange quark affects the single hadron fragmentations of the driving terms, as in their respective integral equations there is a sum over $Q$ as well~[Eq.~\eqref{EQ_SINGEL_FRAG}]. This potentially can have a large effect, as in Sec.~\ref{sec:dfcontri} it was shown that the driving functions give most of the contribution to the favored DFFs. 

The solution of $D^{\pi^+\pi^-}_q$ for $z_1$ fixed at $0.1$ and $0.5$ are shown in Fig.~\ref{fig:dffsquark}. Here, the dashed blue line and the red dot-dashed line represent the results for up quark DFF with and without the strange quark~[denoted by (s) and (no s) in the captions], respectively. Similarly, the purple dotted line and the black solid line represent the results for the down quark DFF with and without the strange quark, respectively.

The shapes of the dihadron fragmentation functions remain the same for both the $u$ and $d$ quark DFFs, with the down quark DFF being the larger in magnitude compared to the up quark DFF for low $z_1$ and vice versa when $z_1$ is increased. The main change is the considerable reduction in the magnitude of DFFs when the strange quark is included, caused by the availability of the kaon emission channels. Thus the inclusion of the strange quark in our model proves to be very important in describing the light quark DFFs. 

\subsection{Study of $D^{\pi^+K^-}_q$}
\label{sec:zfixed}
\begin{figure}[H!t]
\begin{center}
\subfigure[$z_1=0.1$: $z_2\,D^{\pi^+K^-}_q$]{\label{fig:zdudspplkmiz_1eq0_1}
\includegraphics[width=0.45\textwidth]{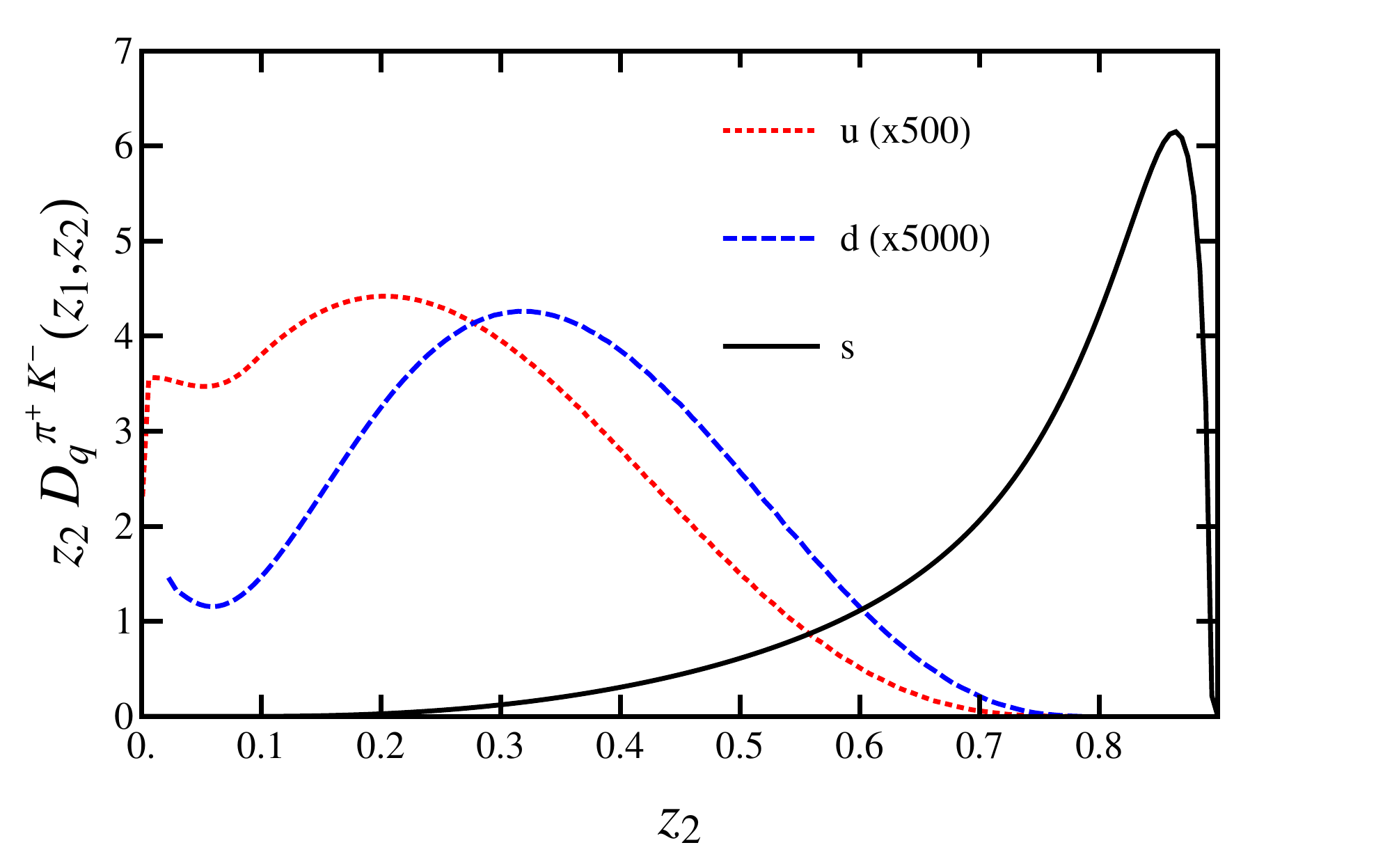}}
\subfigure[$z_2=0.1$: $z_1\,D^{\pi^+K^-}_q$]{\label{fig:zdudspplkmiz_2eq0_1}
\includegraphics[width=0.45\textwidth]{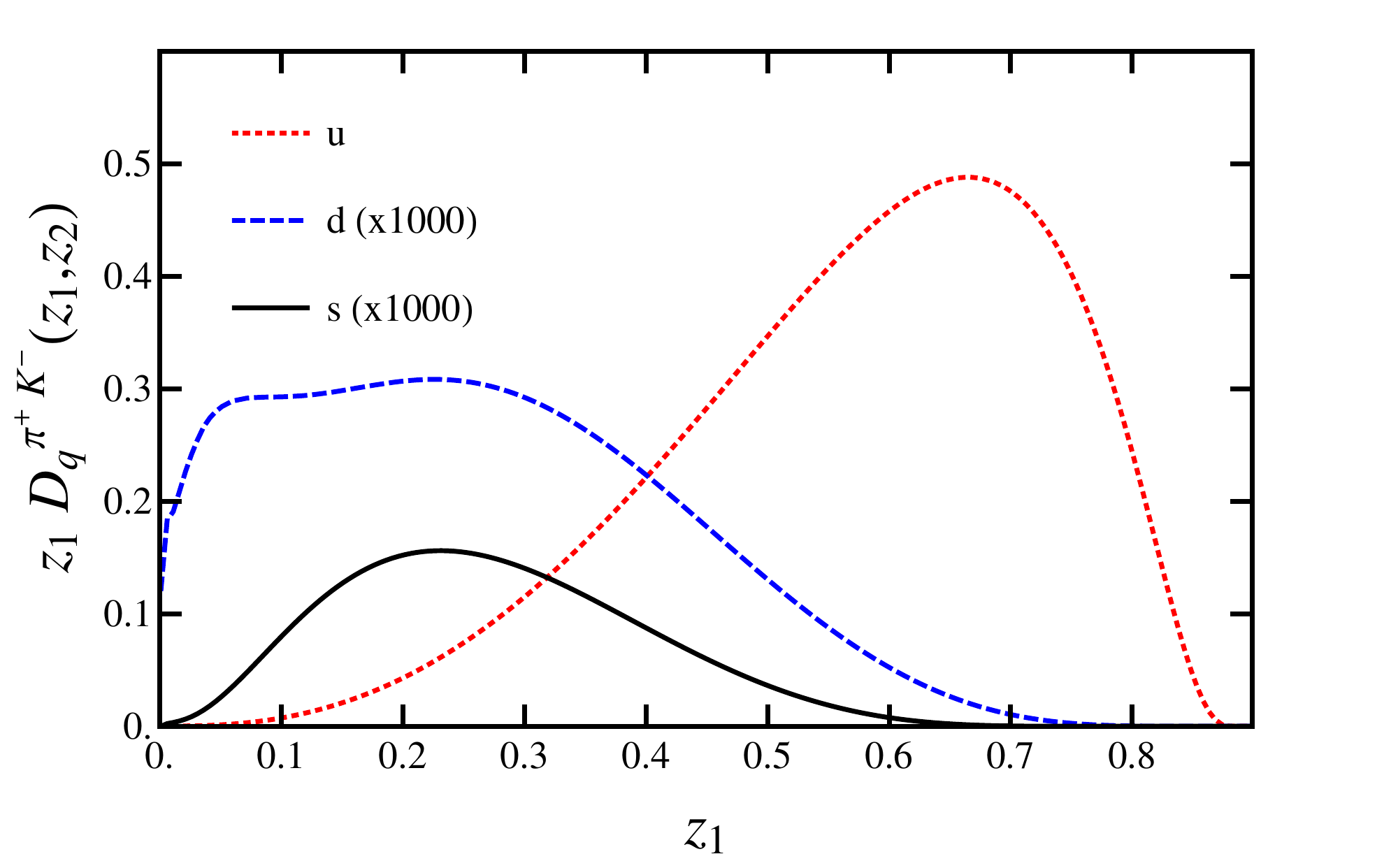}}  
\caption{$\pi^+K^-$ dihadron fragmentation functions for~\subref{fig:zdudspplkmiz_1eq0_1} $z_1=0.1$ and~\subref{fig:zdudspplkmiz_2eq0_1} $z_2=0.1$. The up, down and strange quark DFFs are represented by dotted red lines, dashed blue lines and solid black lines, respectively. The number in the brackets in the legend indicates the scaling factor used in depicting the curve.}
\label{fig:z1z201}
\end{center}
\end{figure}

\begin{figure}[[H!t]
\begin{center}
\subfigure[$z_1=0.5$: $z_2\,D^{\pi^+K^-}_q$]{\label{fig:zdudspplkmiz_1eq0_5}
\includegraphics[width=0.45\textwidth]{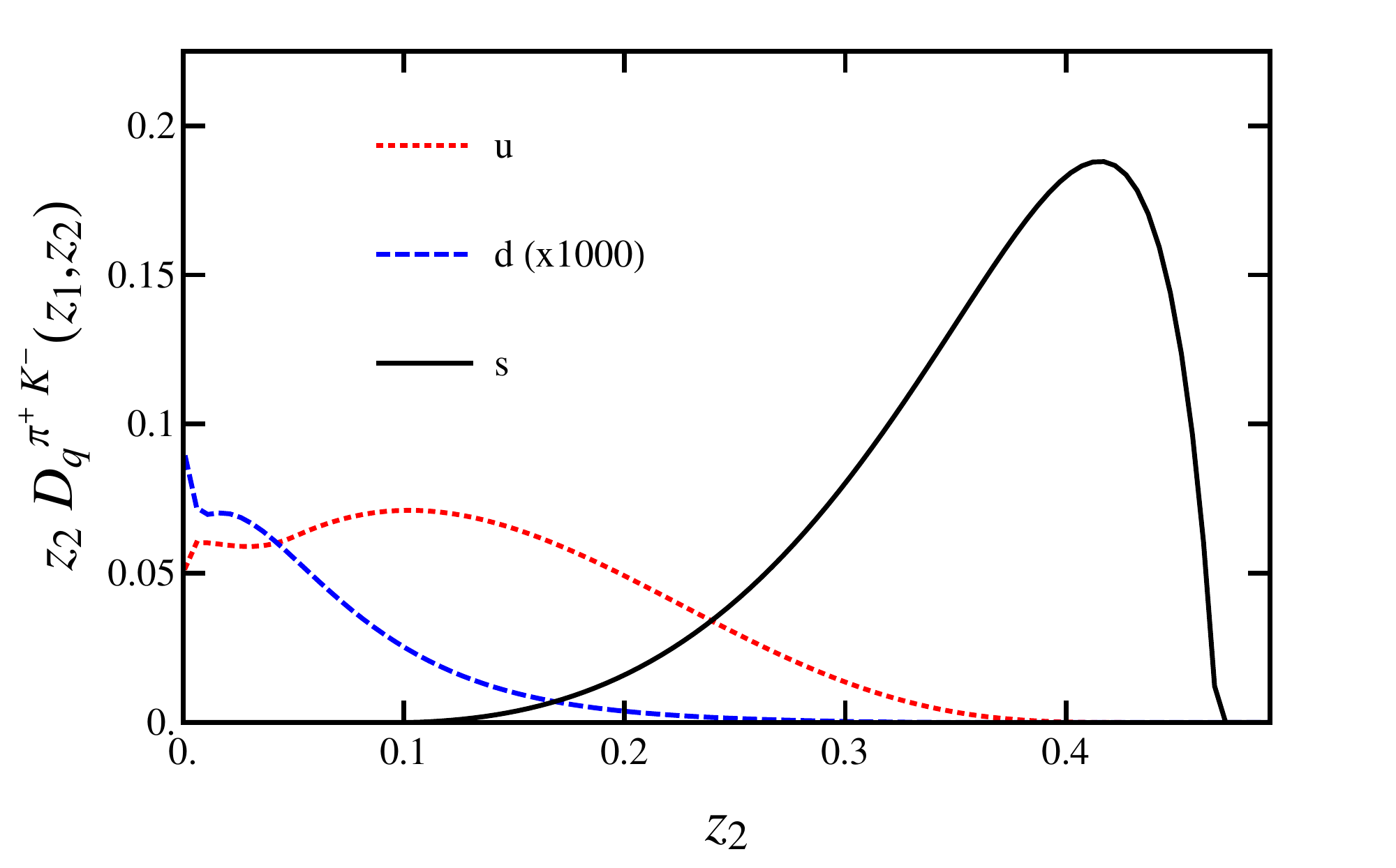}}
\subfigure[$z_2=0.5$: $z_1\,D^{\pi^+K^-}_q$]{\label{fig:zdudspplkmiz_2eq0_5}
\includegraphics[width=0.45\textwidth]{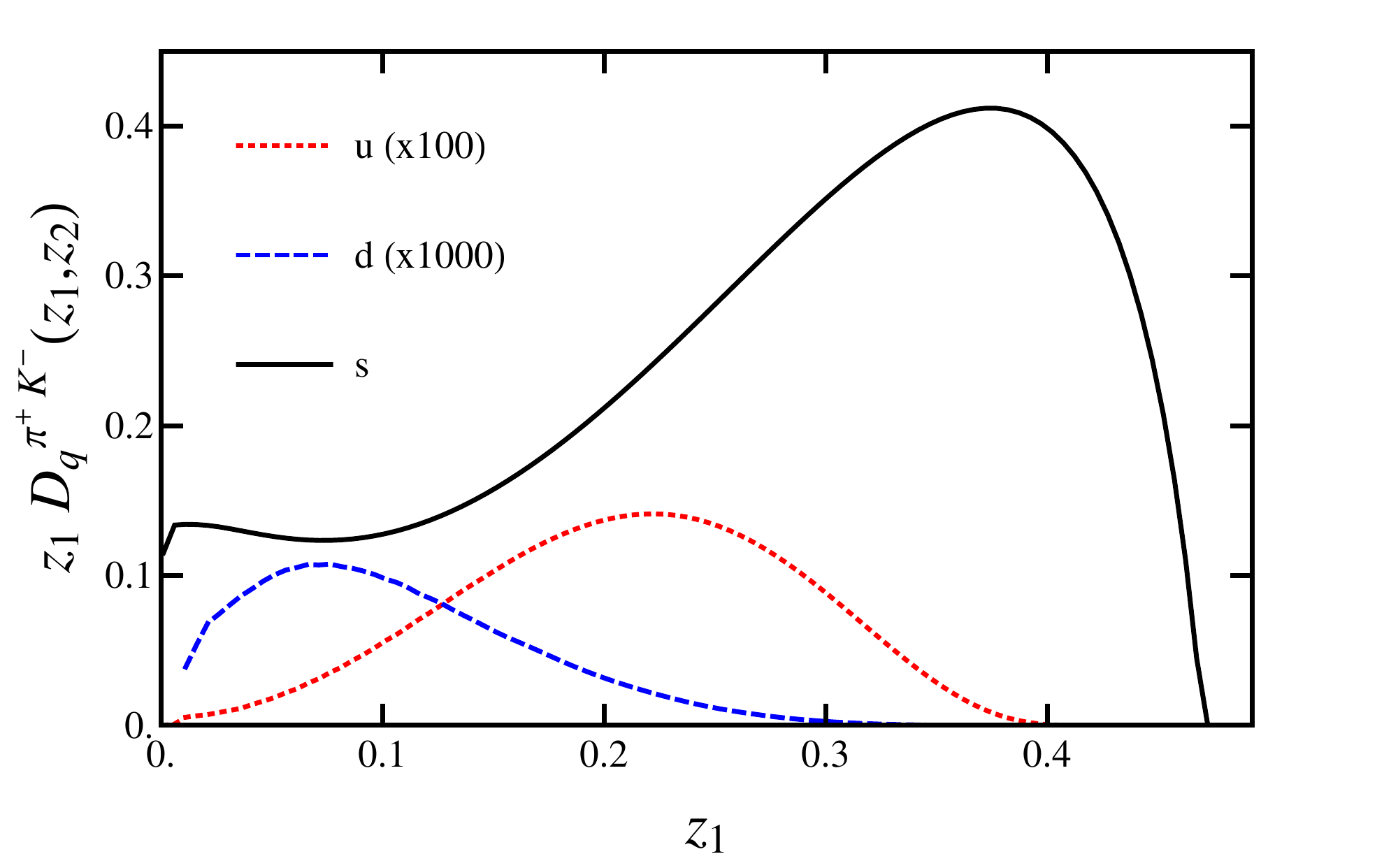}}
\caption{$\pi^+K^-$ dihadron fragmentation functions for~\subref{fig:zdudspplkmiz_1eq0_5} $z_1=0.5$ and~\subref{fig:zdudspplkmiz_2eq0_5} $z_2=0.5$. The up, down and strange quark DFFs are represented by dotted red lines, dashed blue lines and solid black lines, respectively. The number in the brackets in the legend indicates the scaling factor used in depicting the curve.}
\label{fig:z1z205}
\end{center}
\end{figure}

In this section, we examine the plots of $D^{\pi^+K^-}_q$, where either $z_1$ or $z_2$ is fixed. These particular DFFs were chosen since $q\to\pi^+ K^-$ is a favored fragmentation channel to one of the hadrons both for a light and a strange quark $q$. This produces more interesting results to examine than if we had chosen $D^{\pi^+\pi^-}_q$, as those DFFs are symmetric in $q=u$ and $q=d$; thus the DFF for $q=u$ at fixed $z_1$ is the same as that for the DFF for $q=d$ at fixed $z_2$, etc. The results for fixed values of $z_1$ and $z_2$ are shown on the plots in Figs.~\ref{fig:z1z201} and~\ref{fig:z1z205}. The up quark DFFs are represented by dotted red lines, while the down quark DFFs are represented by dashed blue lines and the strange quark DFFs are represented by solid black lines. 

We first examine the DFFs for $z_1=0.1$ [Fig.~\ref{fig:zdudspplkmiz_1eq0_1}] and $z_2=0.1$ [Fig.~\ref{fig:zdudspplkmiz_2eq0_1}]. Since the hadron corresponding to the fixed light-cone momentum fraction only has a small amount of the fragmenting quark's momentum, most of the momentum is attributed to the favored fragmentation channel of the other hadron. For Figs.~\ref{fig:zdudspplkmiz_1eq0_1} and~\ref{fig:zdudspplkmiz_2eq0_1}, this corresponds to the strange and up quark's fragmentations to the $K^-$ and $\pi^+$, respectively. The down quark is unfavored for both hadrons and thus receives very little contribution to its DFF in both plots.

After increasing the fixed value of $z_1$~[Fig.~\ref{fig:zdudspplkmiz_1eq0_5}] and $z_2$~[Fig.~\ref{fig:zdudspplkmiz_2eq0_5}] to $0.5$, the strange quark DFFs are the largest for both. The strange quark is the only initial quark that can produce both hadrons in the first two steps of the cascade, whereas both the up and down quarks require multiple decays to produce both hadrons. The up quark's solutions are the second largest since it can produce the $\pi^+$ hadron in the first emission step, while the down quark's solutions are low for both plots, as it can't produce either of the hadrons in the first emission step.

\section{Evolution of the DFFs}
\label{SEC_EVOL}

The results for the dihadron fragmentation functions in the NJL-jet model presented in Sec.~\ref{sec:results} are all at the model scale of $0.2~\mathrm{ GeV}^2$. The model scale was obtained in Ref.~\cite{Matevosyan:2010hh} such that after NLO evolution the model-calculated $u$ quark valence distribution function in the $\pi^+$ matched those experimentally measured in Refs.~\cite{Sutton:1991ay} and~\cite{Wijesooriya:2005ir}. To compare our results to experiment or results from other models, we need to evolve the dihadron fragmentation functions to an appropriate momentum scale. In Ref.~\cite{deFlorian:2003cg}, the dihadron fragmentation functions evolution equations are derived from factorization of the cross-section for the production of two hadrons in $e^+e^-$ annihilation in the $\overline{\text{MS}}$ factorization scheme. Using JetCalculus, Ref.~\cite{Ceccopieri:2007ip} deduces the evolution equations for DFFs with an explicit dependence on the invariant mass of the hadron pairs, $M_h$. The DFFs with a dependence on the invariant mass are addressed as extended dihadron fragmentation functions~(extDiFF). The extDiFFs are important as they will relate to experimental results that include the dependence on invariant mass spectra.

The leading order (LO) evolution equation for DFF from Ref.~\cite{Ceccopieri:2007ip} is presented in Eq.~(\ref{DFFEVOeq}).
\begin{widetext}
\begin{eqnarray}
\label{DFFEVOeq}
\frac{d}{dlnQ^2} D^{h_1 h_2}_i(z_1,z_2,Q^2) & = & \frac{\alpha_s(Q^2)}{2\pi}\times\int^1_{z_1+z_2} \frac{du}{u^2} D^{h_1 h_2}_j\left(\frac{z_1}{u},\frac{z_2}{u},Q^2\right)P_{ji}(u) \nonumber \\
& & +  \frac{\alpha_s(Q^2)}{2\pi}  \times  \int^{1-z_2}_{z_1} \frac{du}{u(1-u)}D^{h_1}_j\left(\frac{z_1}{u},Q^2\right)D^{h_2}_k\left(\frac{z_2}{1-u},Q^2\right)\hat{P}^i_{jk}(u),
\end{eqnarray}
\end{widetext}
where $Q^2$ is the momentum scale, $\alpha_s(Q^2)$ is the strong coupling at that momentum scale. 

On the left-hand side of Eq.~(\ref{DFFEVOeq}), the rate that the DFFs change with respect to $\ln Q^2$ is represented. The first term on the right-hand side represents the effect of the parton $i$ emitting a parton $j$ with light-cone momentum fraction $u$, with probability $P_{ji}(u)$ that it produces the two detected hadrons, $h_1$ and $h_2$, while the second term represents the effect of two partons, $j$ and $k$, being emitted by $i$ with light-cone momentum fractions $u$ and $1-u$, respectively, with probability $\hat{P}^i_{jk}(u)$, and each of these partons producing one of the detected hadrons.

\begin{figure}[H!t]
\begin{center}
\subfigure[$z_1=0.5$: $z_2\,D^{\pi^+ \pi^-}_u$]{\label{fig:zDupppm4_z1_05}
\includegraphics[width=0.48\textwidth]{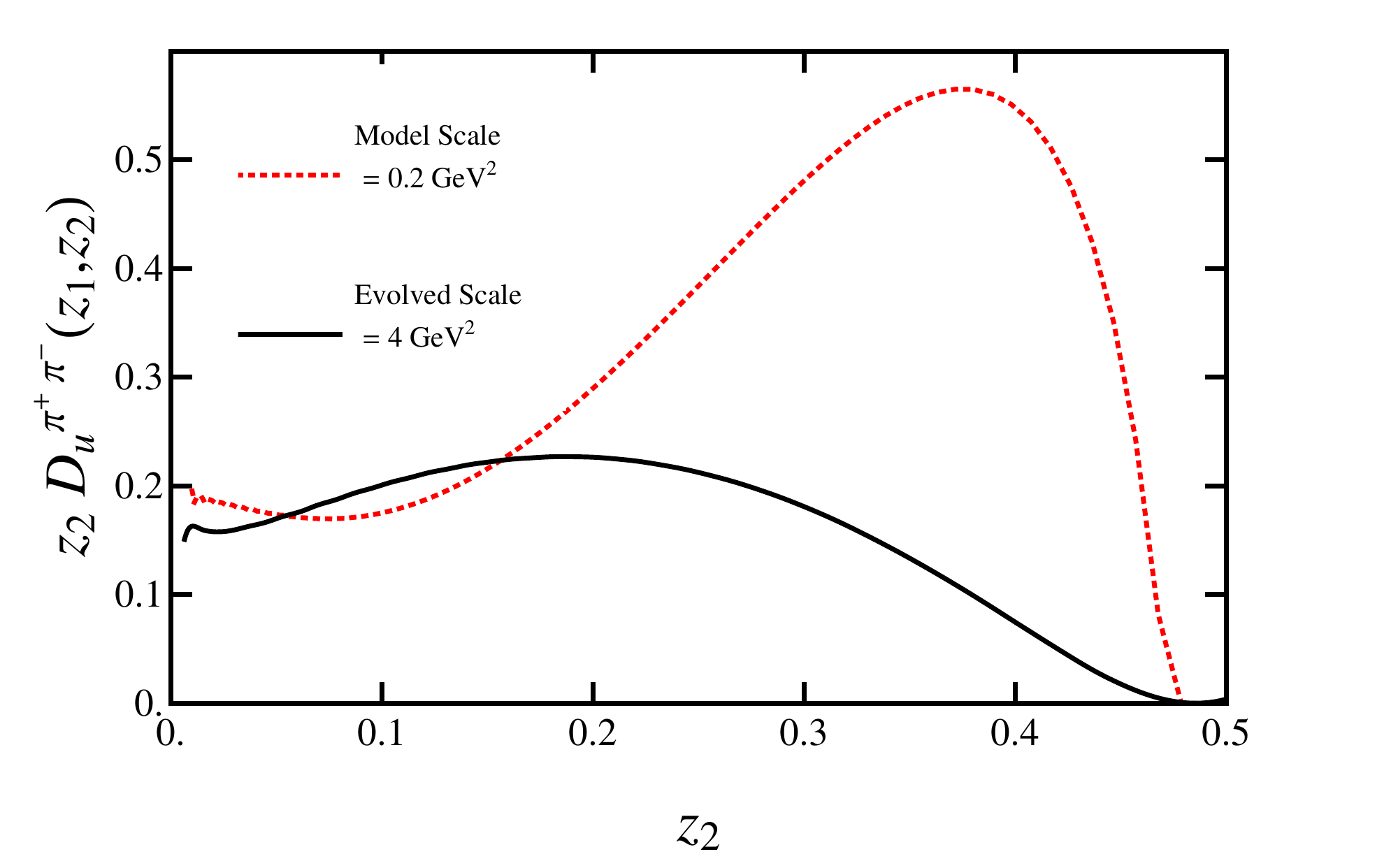}}
\subfigure[$z_2=0.5$: $z_1\,D^{\pi^+ \pi^-}_u$]{\label{fig:zDupppm4_z2_05}
\includegraphics[width=0.48\textwidth]{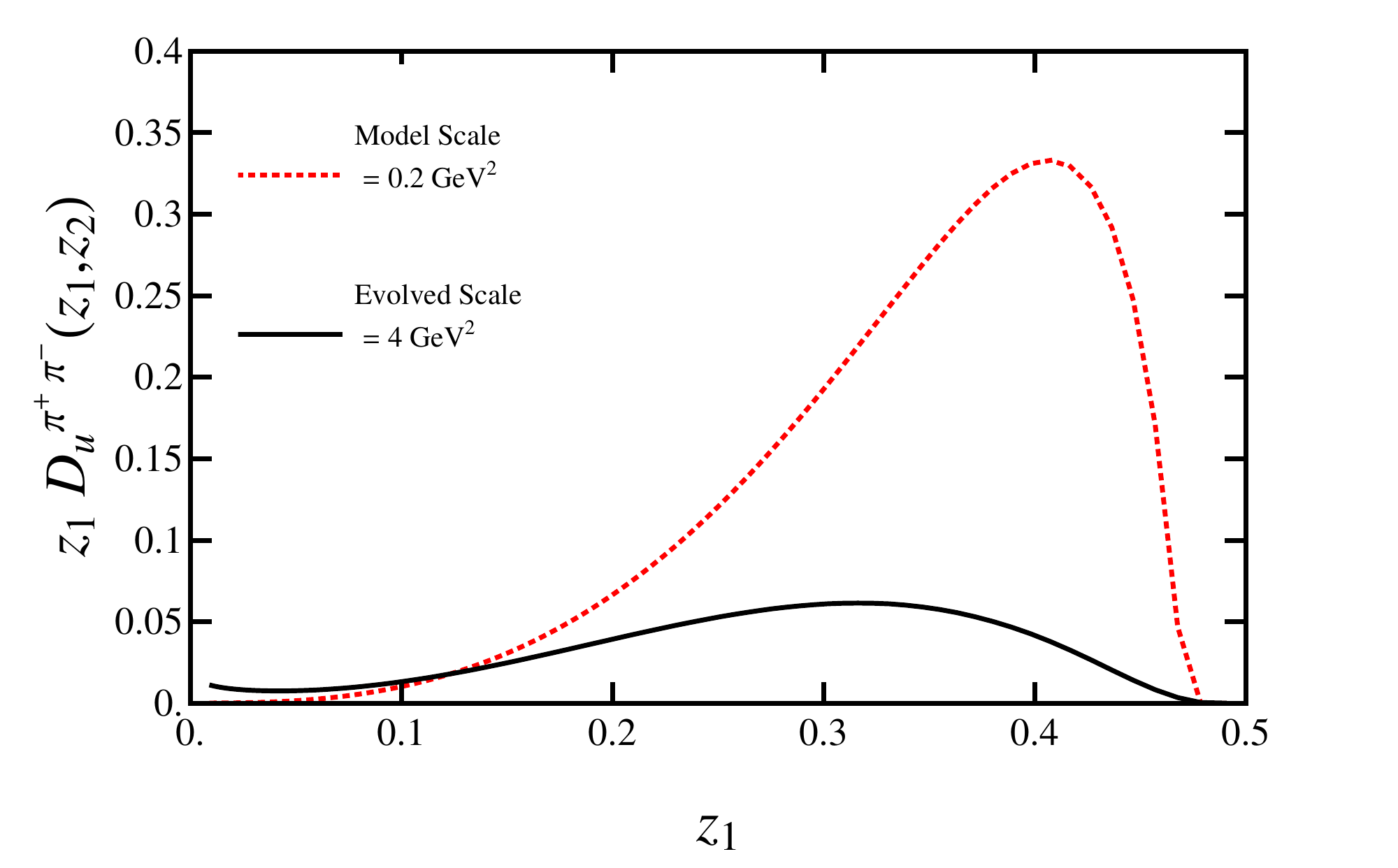}}  
\caption{$u\to \pi^+ \pi^-$ dihadron fragmentation functions are represented as the dotted red lines for the solutions at model scale ($Q_0^2=0.2 \text{ GeV}^2$) and solid black lines for the solutions at the evolved scale ($Q^2=4 \text{ GeV}^2$) for~\subref{fig:zDupppm4_z1_05} $z_1=0.5$ and~\subref{fig:zDupppm4_z2_05} $z_2=0.5$.}
\label{fig:4gevplots}
\end{center}
\end{figure}

We developed a computer code to perform the QCD evolution of the dihadron fragmentation functions according to Eq.~(\ref{DFFEVOeq}), where the DFFs were separated into nonsinglet, singlet and gluon dihadron fragmentation functions. The code is based on the single hadron fragmentation function evolution program by the authors of Refs.~\cite{Miyama:1995bd,Hirai:1996hv,Hirai:1997gb,Hirai:2011si}. The details on the evolution method employed, along with the full set of the results, will be presented in our upcoming paper~\cite{caseyinprep}. As an example, here we present the results for LO evolution of $D_{u}^{\pi^+ \pi^-}(z_1,z_2)$ from our model scale of $0.2~\mathrm{GeV}^2$ to the typical experimental scale of $4~\mathrm{GeV}^2$. The results for the evolved DFFs are presented in Figs.~\ref{fig:zDupppm4_z1_05} and~\ref{fig:zDupppm4_z2_05}, corresponding to the solutions at $z_1=0.5$ and $z_2=0.5$, respectively. The dotted red line represents the solution at the model scale ($0.2~\mathrm{GeV}^2$) and the solid black line represents the solution at the final scale $Q^2=4~\mathrm{GeV}^2$. Both Figs.~\ref{fig:zDupppm4_z1_05} and~\ref{fig:zDupppm4_z2_05} show a shift in the peak of the model results towards the lower $z$ region after the evolution, similar to the single hadron evolution. 

\section{Conclusions and Outlooks}
In this paper we have presented results for dihadron fragmentation functions calculated within the NJL-jet model. DFFs were obtained as numerical solutions of the corresponding integral equations derived using the quark-jet description of the hadronization process. In Sec.~\ref{sec:dfcontri}, we showed that the integral term, that represents the effects of initial undetected hadron emission, has a very small effect on the DFFs, except when the driving function was zero or when $z_1$ was low. For driving functions equal to zero, the corresponding DFFs were generated entirely by the integral term and when $z_1$ was lowered to $0.1$, the relative contribution of the driving function to the DFF was also lowered for most values of $z_2$. For the $u\to \pi^+ \pi^-$ DFF, the peak value of $z_2 D_q^{h_1 h_2}(z_1,z_2)$ at $z_1=0.5$ was almost ten times the peak value at $z_1=0.1$. Because of the lower value of the DFF at low $z_1$, the integral term contribution becomes a more significant part of the DFF, reducing the relative contribution of the driving function. This effect occurs when the fragmenting quark is the favored quark for the hadron that receives a small light-cone momentum fraction, but is unfavored for the hadron that has access to most of the light-cone momentum of the fragmenting quark. One example where this effect is particularly visible is in Fig.~\ref{fig:df_uds_kplkmi_z1eq0_1}, where the $u\to K^+K^-$ DFF is mostly composed of the integral term at low $z_2$ and the driving function at higher $z_2$. In all three results with low $z_1$~(Fig.~\ref{fig:drivfunc01}), the effect is seen for the up quark DFF, which is the favored quark for the hadrons $h_1=\pi^+$ and $h_1=K^+$, but is unfavored for hadrons $h_2=\pi^-$ and $h_2=K^-$.

In Sec.~\ref{sec:strangecontri} we showed that the strange quark's inclusion has a significant impact on the DFFs. The main change to the DFFs when the strange quark is included is a considerable reduction in magnitude, similar to the single hadron fragmentation functions~\cite{Matevosyan:2010hh}, caused by the availability of the kaon emission channels. The comparison plots for the $\pi^+\pi^-$ dihadron fragmentation functions were shown in Fig.~\ref{fig:dffsquark}, where the results are similar for the $\pi^+\pi^+$ and $\pi^+\pi^0$ DFFs.

We examined $D^{\pi^+K^-}_q(z_1,z_2)$, where either $z_1$ or $z_2$ is fixed~(Figs.~\ref{fig:z1z201} and~\ref{fig:z1z205}), in Sec.~\ref{sec:zfixed}. The $\pi^+K^-$ DFFs were chosen as they are favored both for a light quark and a strange quark. At low values of $z_1$ and $z_2$, the strange quark and up quark DFFs, respectively, are dominant. The strange quark DFF is dominant for low $z_1$ because the $\pi^+$ has a small fraction of the light-cone momentum of the initial quark, leaving most of the initial momentum available to the strange quark's favored fragmentation to $K^-$. Similarly, the up quark is dominant for low $z_2$ because the $K^-$ has a small light-cone momentum fraction, allowing the $\pi^+$ to access most of the momentum for its favored fragmentation. Increasing the fixed value of either $z_1$ or $z_2$  to $0.5$ in the $\pi^+K^-$ DFF shows that the strange quark DFF is dominant. This can be easily interpreted within the model, as only the strange quark can produce both the $K^-$ then the $\pi^+$ in the first two steps of the decay chain, whereas the up and down quarks both require at least three steps in the decay chain to produce both hadrons.

Finally, in Sec.~\ref{SEC_EVOL} we discuss the QCD evolution of the dihadron fragmentation functions, which is essential in comparing our model calculations with experimental extractions, as well as Monte Carlo simulations or other analytical results. The evolution equations are presented and the method for the numerical solutions is briefly discussed.  As an example, the results for $D_u^{\pi^+\pi^-}$ presented in Fig.~\ref{fig:4gevplots}, show the significant modification of the DFFs with evolution. The details for solving the evolution equations and the complete set of results for evolved DFFs will be presented in our upcoming paper~\cite{caseyinprep}. A comparison between our results and others will also be presented in that work.

Future work to extend the NJL-jet model for DFFs include the inclusion of hadronic resonances and their decays, as well as the inclusion of the transverse momentum dependence. These have been accomplished in the single hadron fragmentations~\cite{Matevosyan:2011ey,Matevosyan:2011vj} using a Monte Carlo framework. These extensions for the DFFs are certainly possible, but lay beyond the scope of the current work and are left for the future.

\acknowledgements{
This work was supported by the Australian Research Council through Australian Laureate Fellowship FL0992247~(AWT) and the ARC Centre of Excellence for Particle Physics at the Terascale and by the University of Adelaide.}

\bibliographystyle{apsrev}
\bibliography{dihadbib240712}

\end{document}